\documentstyle[12pt,amsfonts]{article}
\parskip=\medskipamount

\newcommand {\cD}{{\cal D}}

\newcommand {\cK}{{\cal K}}
\newcommand {\cL}{{\cal L}}

\newcommand {\cN}{{\cal N}}

\newcommand {\cW}{{\cal W}}

\def\a{\alpha}
\def \bi{\bibitem}

\def\b{\beta}

\def\d{\delta}

\def\f{\phi}
\def\g{\gamma}
\def\G{\Gamma}

\def\k{\kappa}
\def\l{\lambda}
\def\m{\mu}

\def\p{\pi}
\def\q{\theta}

\def\s{\sigma}
\def\t{\tau}

\def\x{\xi}

\def\D{\Delta}
\def\F{\Phi}
\def\J{\Psi}
\def\L{\Lambda}
\def\O{\Omega}

\def\U{\Upsilon}

\newcommand{\ad}{{\dot{\alpha}}}                           
\newcommand{\bd}{{\dot{\beta}}}                            
\newcommand{\ve}{\varepsilon}                            
\newcommand{\cDB}{{\bar \cD}}                            
\newcommand{\gd}{{\dot{\gamma}}}

\newcommand{\pa}{\partial}                           
\newcommand{\hf}{\frac12}

\newcommand{\vf}{\varphi}
\newcommand{\sect}[1]{\setcounter{equation}{0}\section{#1}}

\newcommand{\be}{\begin{equation}}
\newcommand{\ee}{\end{equation}}
\newcommand{\bea}{\begin{eqnarray}}
\newcommand{\eea}{\end{eqnarray}}
\newcommand{\non}{\nonumber}

\def \intss{\int\!\!{\rm d}^8z}
\def \intssc{\int\!\!{\rm d}^6z}

\def \ERc{\frac{E^{-1}}{R}}
\def \ERac{\frac{E^{-1}}{\bar R}}
\def \Dsqc{(\cD^2 - 4 {\bar R})}
\def \Dsqac{(\cDB^2 - 4 R)}
\newcommand{\dsR}{{\Bbb R}}

\begin{document}

\begin{titlepage}

\begin{flushright}
hep-th/0212039 \\
December 2002\\
Revised version: February 2003
\end{flushright}
\vspace{5mm}

\begin{center}
{\Large\bf  Nonlinear
Self-Duality and Supergravity}
\end{center}
\vspace{3mm}

\begin{center}

{\large Sergei M. Kuzenko and Shane A. McCarthy}
\vspace{2mm}

\footnotesize{
{\it School of Physics, The University of Western Australia,\\
35 Stirling Highway, Crawley W.A. 6009, Australia}} \\
{\tt kuzenko@cyllene.uwa.edu.au}~,
{\tt shane@physics.uwa.edu.au}
\vspace{2mm}

\end{center}
\vspace{5mm}

\begin{abstract}
\baselineskip=14pt
\noindent
The concept of self-dual supersymmetric nonlinear electrodynamics is
generalized to a curved superspace of $\cN=1$ supergravity, for both
the old minimal and the new minimal versions of $\cN=1$ supergravity.
We derive the self-duality equation, which has to be satisfied by
the action functional of any U(1) duality invariant model of
a massless vector multiplet, and construct a family of self-dual
nonlinear models. This family includes a curved superspace extension
of the $\cN=1$ super Born-Infeld action. The supercurrent and supertrace
in such models are proved to be duality invariant. The most interesting
and unexpected result is that the requirement of nonlinear self-duality
yields nontrivial couplings of the vector multiplet to K\"ahler sigma 
models.
We explicitly derive the couplings to general K\"ahler sigma models
in the case when the matter chiral multiplets are inert under
the duality rotations, and more specifically to the dilaton-axion
chiral multiplet when the group of duality rotations is enhanced
to SL$(2,{\Bbb R})$.
\end{abstract}
\vfill
\end{titlepage}

\newpage
\setcounter{page}{1}
\renewcommand{\thefootnote}{\arabic{footnote}}
\setcounter{footnote}{0}

\sect{Introduction}

In 1935, Schr\"odinger \cite{Sch} showed that the nonlinear
electrodynamics of Born and Infeld \cite{BI},
proposed as a new fundamental theory of the electromagnetic
field, possessed a remarkable
property -- invariance under U(1) duality rotations.
Although the great expectations, which originally led the authors
of \cite{BI} to put forward their model, never came true,
the Born-Infeld action has re-appeared in the spotlight since the 1980's
as a low energy effective action in string theory \cite{FT,L}.
Along with patterns of duality in extended supergravity \cite{FSZ,CJ},
this motivated Gaillard and Zumino, Gibbons and others
\cite{GZ1,Z,GR1,GR2,GZ2,GZ3,list}
to develop a general theory of (nonlinear) self-duality in four
and higher space-time dimensions for non-supersymmetric 
theories\footnote{Properties of nonlinear electrodynamics 
in curved space from the viewpoint of dualities 
were studied in \cite{GH}.}.
Extension to 4D $\cN=1, \,2$ globally supersymmetric theories
was given in \cite{KT1,KT2}.
As a final step in developing the formalism,
it was also shown in \cite{IZ} how to re-formulate
the requirement of nonlinear self-duality (i.e. a nonlinear equation
which the action functional has to satisfy in order for the theory to be
duality invariant) as a condition of manifest
invariance of the interaction; in a sense, this is
a nice extension of Schr\"odinger's ideas \cite{Sch}.

There exist deep yet mysterious connections between nonlinear 
self-duality
and supersymmetry and here we give three examples. First,
in the case of partial spontaneous
supersymmetry breakdown $\cN=2 \to \cN=1$,
the Maxwell-Goldstone multiplet \cite{BG1,RT}
(coinciding with the $\cN=1$ supersymmetric Born-Infeld
action \cite{CF}) and the tensor Goldstone multiplet
\cite{BG2,RT} were shown  in \cite{KT1,KT2} to be
self-dual, i.e. invariant under U(1) duality rotations.
Our second example concerns
partial supersymmetry breakdown $\cN=4 \to \cN=2$.
To construct a Maxwell-Goldstone multiplet action for such
a scenario -- the $\cN=2$ supersymmetric Born-Infeld
action\footnote{See \cite{Ket} for earlier attempts
to construct an $\cN=2$ supersymmetric version of the
Born-Infeld action.} --
it was suggested in \cite{KT2} to look for an
$\cN=2$ vector multiplet action which should be
(i) self-dual; (ii) invariant under a nonlinearly realized
central charge bosonic symmetry. These requirements
turn out to allow one
to restore the Goldstone multiplet action uniquely
to any fixed order in powers of chiral superfield strength $\cW$;
this was carried out in \cite{KT2} up to order $\cW^{10}$.
Recently, there has been considerable progress in developing
the formalism of nonlinear realizations to describe
the partial SUSY breaking $\cN=4 \to \cN=2$ \cite{BIK}.
So far, the authors of \cite{BIK} have reproduced the action
obtained in \cite{KT2}. Finally, we should mention that
the (Coulomb branch) low energy effective action of the
$\cN=4$ supersymmetric Yang-Mills theory is conjectured to be
invariant under U(1) duality rotations \cite{KT1}
(a weaker form of self-duality of the effective action
was proposed in \cite{GKPR}).

The above features provide enough evidence for considering
supersymmetric self-dual systems to be quite interesting and their
properties worth studying. In the present note,
self-dual nonlinear supersymmetric electrodynamics \cite{KT1,KT2}
is coupled to $\cN=1$ supergravity in the presence of nonlinear
K\"ahler sigma models. To describe $\cN=1$ supergravity,
we separately consider its two off-shell
realizations\footnote{The non-minimal version of $\cN=1$
supergravity \cite{non-min} does not lead to interesting
matter couplings, see e.g. \cite{BK}.}:
the old minimal formulation \cite{WZ-old,old} and the new minimal
formulation \cite{new}. By now, 4D $\cN=1$ superfield supergravity
is a subject of several textbooks \cite{GGRS,WB,BK}
to which the interested reader is referred for further
details and references.

The structure of the paper is as follows.
Working with the old minimal version of $\cN=1$ supergravity,
in  section \ref{sect:sd} we derive the self-duality
equation as the condition for a vector multiplet model to be
invariant under U(1) duality rotations. We then
demonstrate self-duality under a superfield Legendre
transformation, introduce a family of self-dual nonlinear
models and argue that the supercurrent and supertrace in such
models are duality invariant. To a large extent, these
results are just a minimal curved superspace extension of the globally
supersymmetric results presented in \cite{KT2}.
However, the game will become more interesting
when we turn to the couplings to new minimal supergravity
and K\"ahler sigma models.
Section \ref{sect:matter} is devoted to the coupling
of self-dual nonlinear supersymmetric electrodynamics
to the dilaton-axion multiplet in curved superspace;
this is of interest from the point of view of string theory.

\sect{Electromagnetic duality rotations in curved
superspace}\label{sect:sd}

We follow the notation\footnote{
In particular, $z^M = (x^m, \q^\m , {\bar \q}_{\dot \m})$ are the
coordinates of $\cN=1$ curved superspace,
${\rm d}^8z = {\rm d}^4 x\, {\rm d}^2 \q \,{\rm d}^2 {\bar \q}$ is
the full flat superspace measure, and
${\rm d}^6z = {\rm d}^4 x\, {\rm d}^2 \q$ is the measure in the
chiral subspace.}
and $\cN=1$ supergravity conventions of \cite{BK}.
Unless otherwise stated we work with the old
minimal formulation of $\cN=1$ supergravity.
The superspace geometry is described by covariant derivatives
\bea
\cD_A &=& (\cD_a , \cD_\a ,\cDB^\ad ) = E_A + \O_A~, \non \\
E_A &=& E_A{}^M \pa_M  ~, \qquad
\O_A = \hf\,\O_A{}^{bc} M_{bc}
= \O_A{}^{\b \g} M_{\b \g}
+\O_A{}^{\bd \gd} {\bar M}_{\bd \gd} ~,
\eea
with $E_A^{~M}$ the supervielbein,
$\O_A$ the Lorentz superconnection
and $M_{bc} \Leftrightarrow ( M_{\b\g}, {\bar M}_{\bd \gd})$
the Lorentz generators.
The covariant derivatives obey the following algebra:
\bea
&& {} \qquad \{ \cD_\a , {\bar \cD}_\ad \} = -2{\rm i} \cD_{\a \ad} ~, 
\non \\
\{\cD_\a, \cD_\b \} &=& -4{\bar R} M_{\a \b}~, \qquad
\{ {\bar \cD}_\ad, {\bar \cD}_\bd \} = - 4R {\bar M}_{\ad \bd}~,  \\
\left[ { \bar \cD}_{\ad} , \cD_{ \b \bd } \right]
     & = & -{\rm i}{\ve}_{\ad \bd}
\Big(R\,\cD_\b + G_\b{}^{\dot{\g}}  \cDB_{\dot{\g}}
-(\cDB^\gd G_\b{}^{\dot{\d}})
{\bar M}_{\gd \dot{\d}}
+2W_\b{}^{\g \d}
M_{\g \d} \Big)
- {\rm i} (\cD_\b R)  {\bar M}_{\ad \bd}~,  \non  \\
\left[ \cD_{\a} , \cD_{ \b \bd } \right]
     & = & \phantom{-}{\rm i}{\ve}_{\a \b}
\Big({\bar R}\,\cDB_\bd + G^\g{}_\bd \cD_\g
- (\cD^\g G^\d{}_\bd)  M_{\g \d}
+2{\bar W}_\bd{}^{\gd \dot{\d}}
{\bar M}_{\gd \dot{\d} }  \Big)
+ {\rm i} (\cDB_\bd {\bar R})  M_{\a \b}~,  \non
\eea
where the tensors $R$, $G_a = {\bar G}_a$ and
$W_{\a \b \g} = W_{(\a \b\g)}$ satisfy the Bianchi identities
\be
\cDB_\ad R= \cDB_\ad W_{\a \b \g} = 0~, \quad
\cDB^\gd G_{\a \gd} = \cD_\a R~, \quad
\cD^\g W_{\a \b \g} = {\rm i} \,\cD_{(\a }{}^\gd G_{\b) \gd}~.
\ee
Modulo purely gauge degrees of freedom, all geometric objects --
the supervielbein and the superconnection -- can be expressed
in terms of three unconstrained superfields
(known as the prepotentials of old minimal supergravity):
gravitational superfield $H^m = {\bar H}^m$,
chiral compensator $\vf$  (${\bar E}_\ad \vf =0$)
and its conjugate $\bar \vf$.
The old minimal supergravity action is
\be
S_{\rm SG,old} = - 3
\intss\,E^{-1}~, \qquad \quad
E = {\rm Ber} (E_A{}^{M})~,
\label{eq:omsg}
\ee
with the gravitational coupling constant being set equal to one.

In what follows, to simplify notation, we introduce
\be
A \cdot B = \intss\, \ERc\, A^\a B_\a~, \qquad
{\bar A} \cdot {\bar B} =\intss\, \ERac\,
{\bar A}_\ad {\bar B}^\ad~,
\ee
with $A_\a$ and $B_\a$ covariantly chiral spinor superfields,
$\cDB_\ad A_\a = \cDB_\ad B_\a =0$
(a similar notation will also be used for chiral scalars).
We often make use of the relation
\bea
\intss\,E^{-1}\,\cL &=& -\frac14 \intss\, \ERc\, \Dsqac \cL \non \\
&=& -\frac14 \intssc\, \vf^3 \,
\Dsqac \cL ~,
\label{chiralrule}
\eea
where the equality in the last line takes place in the so-called
chiral representation (see \cite{GGRS,BK} for more details).
This result is especially simple in the chiral case,
$\cL = \cL_{\rm c} / R$, with $\cL_{\rm c}$ a covariantly chiral
scalar, $\cDB_\ad \cL_{\rm c} = 0$.

\subsection{Self-duality equation}
Consider a model of a single Abelian $\cN = 1$
vector multiplet in curved superspace as generated by the action
$S[W , {\bar W}]$. The covariantly (anti) chiral spinor superfield 
strengths
${\bar W}_\ad$ and $W_\a$,
\be
W_\a = -\frac{1}{4}\, \Dsqac \cD_\a \, V~, \qquad \quad
{\bar W}_\ad = -\frac{1}{4}\, \Dsqc \cDB_\ad \, V ~,
\label{eq:w-bar-w}
\ee
are defined in terms of a real unconstrained prepotential $V$.
Consequently, the strengths are constrained superfields satisfying
the Bianchi identity
\be
\cD^\a W_\a ~=~ \cDB_\ad {\bar W}^\ad~.
\label{eq:bianchi}
\ee

Suppose that $S[W , {\bar W}] \equiv S [v]$ can be unambiguously
defined\footnote{As indicated in \cite{KT1,KT2},
this is always possible if $S[W , {\bar W}]$
does not involve the combination $\cD^\a W_\a $ as an
independent variable.} as a functional of {\it unconstrained}
(anti) chiral superfields ${\bar W}_\ad$ and $W_\a$.
Then, one can define {\it covariantly (anti) chiral} superfields
${\bar M}_\ad$ and $M_\a$  as
\be
{\rm i}\,M_\a \,[v]\equiv 2\, \frac{\d }{\d W^\a}\,S[v]
~, \qquad \quad
- {\rm i}\,{\bar M}^\ad \,[v]\equiv 2\,
\frac{\d }{\d {\bar W}_\ad}\, S[v] ~,
\label{eq:Mdefinition}
\ee
with the functional derivatives defined by
\bea
\d S &=&
\d W \cdot \frac{\d S}{\d W} +
\d {\bar W} \cdot \frac{\d S}{\d {\bar W}} ~.
\eea
The vector multiplet equation of motion following from the action
$S[W,\bar{W}]$ reads
\be
\cD^\a M_\a ~=~ \cDB_\ad {\bar M}^\ad~.
\label{eq:eom}
\ee
Since the Bianchi identity (\ref{eq:bianchi}) and the equation of
motion (\ref{eq:eom}) have the same functional form, one may
consider U(1) duality rotations
\bea
    \left( \begin{array}{c}  M'_\a \,[v'] \\ W'_\a  \end{array} \right)
~=~  \left( \begin{array}{cr} \cos \t ~& ~
-\sin \t \\ \sin \t ~ &  ~\cos \t \end{array} \right) \;
\left( \begin{array}{c}  M_\a \, [v] \\ W_\a  \end{array} \right) ~,
\label{eq:U(1)transformation}
\eea
where $M'$ should be
\be
{\rm i}\,M'_\a \,[v']= 2\, \frac{\d }{\d W'^\a}\,S[v']~.
\ee
Following the method described in \cite{KT1,KT2}, the condition that
$S[W , {\bar W}]$ be self-dual is equivalent to the reality
condition
\be
{\rm Im} \Big( W \cdot W ~+~ M \cdot M \Big) ~=~0~.
\label{eq:duality}
\ee
Any solution, $S[W , {\bar W}]$, of this {\it self-duality equation}
generates a U(1) duality invariant supersymmetric electrodynamics
coupled to old minimal supergravity.
The self-duality equation can be shown to be equivalent to the
following invariance condition under U(1) duality rotations
(\ref{eq:U(1)transformation})
\be
S[v'] - \frac{\rm i}{4} \Big( W' \cdot M' [v']
-{\bar W}' \cdot {\bar M}' [v'] \Big)
= S[v] - \frac{\rm i}{4} \Big( W \cdot M [v]
-{\bar W} \cdot {\bar M} [v] \Big)~.
\label{inv-con}
\ee

\subsection{Invariance under superfield Legendre transformation}
One of the nice properties of all models of
self-dual electrodynamics is
invariance under Legendre transformation \cite{GZ3}.
It was shown in \cite{KT1} that this property
also holds for any globally supersymmetric model
of the massless vector multiplet that is invariant under U(1)
duality rotations.
We will demonstrate that this property also naturally
extends to curved superspace.

Consider a massless vector multiplet model in curved superspace
described by the action $ S[{W},{\bar W} ] $.
The Legendre transformation is defined by introducing an auxiliary 
action
\be
S[\cW , {\bar \cW}, W_{\rm D}, {\bar W}_{\rm D}] =
S[\cW,{\bar \cW} ] - \frac{\rm i}{2}\,
(\cW \cdot W_{\rm D} - {\bar \cW} \cdot {\bar W}_{\rm D})~,
\label{eq:legendre}
\ee
where $\cW_\a$ is now an unconstrained covariantly chiral spinor 
superfield,
and $W_{{\rm D}\, \a}$ the dual field strength
\be
W_{{\rm D}\,\a} = -\frac14\, \Dsqac \cD_\a \, V_{\rm D}~,
\qquad \quad
{\bar W}_{{\rm D}\, \ad }= -\frac14\,\Dsqc {\bar \cD}_\ad \,
V_{\rm D} ~,
\ee
with the Lagrange multiplier $V_{\rm D}$ a real scalar superfield.
This model is equivalent to the original model, since upon elimination 
of
$W_{\rm D}$ by its equation of motion we regain $S[\cW , {\bar \cW}]$,
with the condition that $\cW$ satisfy the Bianchi identity 
(\ref{eq:bianchi}).
However, we may instead eliminate $\cW$ by its equation of motion, in
which case we obtain a dual action
$S_{\rm D}[W_{\rm D} , {\bar W}_{\rm D}]$, the Legendre transform of
$S[W , {\bar W}]$.
Now, let $ S[{W},{\bar W} ] $ be a solution of the self-duality
equation (\ref{eq:duality}), and hence it satisfies
the invariance condition (\ref{inv-con}) under
arbitrary U(1) duality rotations (\ref{eq:U(1)transformation}).
Then, eq. (\ref{inv-con})  for a finite duality rotation
$\t =\p/2$ is easily seen to be equivalent to
the statement that $S_{\rm D} = S$.

\subsection{Family of self-dual models}
Extending the globally supersymmetric results of \cite{KT1},
we now  present a family of $\cN=1$ supersymmetric self-dual models
with actions of the general form
\be
S [W, {\bar W}] =
\frac14\, \intss\, \ERc\, W^2 +
\frac14\, \intss\, \ERac\, {\bar W}^2
+  \frac14\, \intss\, E^{-1} \, W^2\, {\bar W}^2\,
\L(u, {\bar u})~,
\label{eq:family action}
\ee
where $\L(u,\bar u )$ is a real analytic function of the complex  
variable
\be
u ~ \equiv ~ \frac{1}{8} \Dsqc\, W^2~.
\label{u}
\ee
The condition of self-duality
(\ref{eq:duality}) on the model (\ref{eq:family action})
turns out to be equivalent to a differential equation
which the interaction $\L(u, {\bar u})$ has
to satisfy. This  equation is
\be
{\rm Im} \,\Big\{ \G
- \bar{u}\, \G^2
\Big\} = 0~, \qquad \quad
\G  = \frac{\pa (u \, \L) }{\pa u}~.
\label{eq:differential}
\ee

As an important example, consider a minimal curved superspace
extension\footnote{Such a curved superspace action
was discussed in \cite{GK}.}
of the $\cN=1$ supersymmetric Born-Infeld action
\cite{CF,BG1,RT}
\be
S_{\rm BI} =  \frac{1}{4} \intss\, \ERc\, X ~+~
\frac{1}{4} \intss\, \ERac\, {\bar  X}~,
\label{eq:borninfeld1}
\ee
where the covariantly  chiral scalar $X $ is a functional of $W_\a$
and ${\bar W}_\ad$ defined by the nonlinear constraint
\be
X ~+~ \frac{1}{4} \,  X\, \Dsqac\,
{\bar X}  ~=~ W^2~.
\label{eq:constraint}
\ee
${}$Following \cite{BG1}, this can be shown to be equivalent to
\be
S_{\rm BI} =
\frac14\, \intss\, \ERc\, W^2 +
\frac14\, \intss\, \ERac\, {\bar W}^2
+   \intss\, E^{-1} \frac{W^2\,{\bar W}^2  }
{ 1 + \hf\, A \, + \sqrt{1 + A +\frac{1}{4} \,B^2} }~,
\label{eq:born-infeld2}
\ee
where
\be
A = 4(u + \bar u)~, \qquad \quad
B = 4(u - \bar u )~.
\ee
This action is of the form (\ref{eq:family action}),
and one can readily check that
the differential equation (\ref{eq:differential}) is satisfied.
Therefore, the minimal curved superspace extension
of the $\cN=1$ super Born-Infeld action is self-dual.

\subsection{Duality invariance of the supercurrent and supertrace}

In the bosonic case, self-dual models have the important
property that the energy-momentum tensor is invariant under  U(1)
duality rotations \cite{Sch,GZ1,GR1,GR2,GZ3}.
It is natural to ask whether this property extends
to the supersymmetric case, the superfield generalization of the
energy-momentum tensor being the supercurrent $T_a = {\bar T}_a$ and
supertrace $T$, $\cDB_\ad T=0$. These are defined in terms of
covariantized variational derivatives with respect to the supergravity
prepotentials (see \cite{GGRS,BK} for more details),
\be
T_a = \frac{\D S}{\D H^a} ~,
\qquad \quad
T = \frac{\D S}{\D \vf} ~,
\ee
and satisfy the conservation equation
\be
\cDB^{\ad}T_{\a \ad} = -\frac{2}{3}\,\cD_\a T~,
\label{eq:conservation}
\ee
when the matter superfields are put on the mass shell.

Gaillard and Zumino \cite{GZ1,GZ3} developed an elegant,
model-independent proof of the fact that the energy-momentum tensor
of any self-dual bosonic system is invariant under U(1) duality 
rotations.
It is not quite trivial however to generalize this proof
to the supersymmetric case, and this is why we will follow
a brute-force approach, similar to \cite{Sch,GR1,GR2},
and directly check duality invariance of the supercurrent and 
supertrace.

Let us first turn to the supertrace.
${}$For the model (\ref{eq:family action}),  it is
\be
T = \frac{1}{8}\,W^2 \Dsqac\!\left[ {\bar W}^2
\Big(\G +{\bar \G}
- \L \Big)\right]~,
\label{eq:supertrace}
\ee
with $\G$ defined in (\ref{eq:differential}).
Consider an infinitesimal duality rotation
$\d W_\a = \t M_\a~,~\d M_\a = -\t W_\a$, where
\be
{\rm i} \,M_\a = W_\a \left\{ 1 - \frac{1}{4} \Dsqac\! \,
\Big[{\bar W}^2 \Big(\L + \frac{1}{8} \Dsqc
\Big(W^2 \, \frac{\pa \L }{\pa u} \Big) \Big) \Big] \right\}~.
\ee
It is an instructive exercise to show that $\d T$ vanishes
for $\L \neq 0$ only if the self-duality equation 
(\ref{eq:differential})
is taken into account. Now, the conservation equation
(\ref{eq:conservation}) is to be satisfied both before and after
applying the duality rotation. Since $T$ is duality invariant,
the left hand side of (\ref{eq:conservation}) should  also be invariant.
This essentially implies  duality invariance of the supercurrent.

Turning now to the supercurrent, with the use of the techniques 
described in \cite{BK},  we find
\bea
\label{eq:supercurrent}
T_{\a \ad} ~&=&~ {\rm i} M_\a {\bar W}_\ad - {\rm i} W_\a {\bar M}_\ad
~-~ \frac{\rm i}{4} \cD_{\a \ad}\!\left(W^2 {\bar W}^2
\left(\G - {\bar \G} \right)\right) \non \\
&-& \frac16 G_{\a \ad} W^2 {\bar W}^2\!\left(\G +{\bar \G} - \L \right)
~-~ \frac{1}{24} \left[ \cD_\a , \cDB_\ad \right]
\!\left(W^2 {\bar W}^2\!\left(\G +{\bar \G} - \L \right) \right)
\non \\
&-& \frac{\rm i}{4} ( W^2 \stackrel{\longleftrightarrow}{\cD_{\a \ad}}
  {\bar W}^2 ) \L
~-~ \frac{\rm i}{4} W^2 {\bar W}^2 (\cD_{\a \ad} u)
\frac{\pa \L }{\pa u}
~+~ \frac{\rm i}{4} W^2 {\bar W}^2 (\cD_{\a \ad} {\bar u})
\frac{\pa \L }{\pa {\bar u}} \\
&+& \frac{\rm i}{16} (\cD_{\a \ad} W^2) {\bar W}^2
\Dsqc \Big(W^2 \, \frac{\pa \L }{\pa u} \Big)
~-~ \frac{\rm i}{16} W^2 (\cD_{\a \ad} {\bar W}^2)
\Dsqac \Big({\bar W}^2 \, \frac{\pa \L }{\pa {\bar u}} \Big)~. \non
\eea
Off the mass shell, the variational derivative $\D S/\D H$
can be shown to include the extra (gauge non-invariant) term
\be
\frac{\rm i}{4} (\cD^\b M_\b - \cDB_\bd {\bar M}^\bd )
\left[ \cD_\a , \cDB_\ad \right] V~,
\ee
which involves the naked prepotential $V$ 
and therefore does not allow a naive generalization of
the Gaillard-Zumino proof \cite{GZ1,GZ3} to superspace.

A tedious calculation is required to explicitly show that 
(i) the conservation equation (\ref{eq:conservation}) is indeed 
satisfied; and (ii) 
the supercurrent (\ref{eq:supercurrent}) is duality invariant.
When performing these calculations, the following equivalent expression
for the supercurrent is often easier to work with:
\bea
T_{\a \ad} ~&=&~ {\rm i} M_\a {\bar W}_\ad - {\rm i} W_\a {\bar M}_\ad
~+~ \frac{\rm i}{4} \cD_{\a \ad}\!\left(W^2 {\bar W}^2
\!\left(\G + {\bar \G} - \L\right)\right) \non \\
&-& \frac16 G_{\a \ad} W^2 {\bar W}^2 \left(\G +{\bar \G} - \L \right)
~-~ \frac{1}{24} \left[ \cD_\a , \cDB_\ad \right]
\!\left(W^2 {\bar W}^2\!\left(\G +{\bar \G} - \L \right) \right)
\non \\
&-& \frac{\rm i}{2} W^2 {\cD}_{\a \ad}\!\left[{\bar W}^2\!\left( \L
+ \frac18 \Dsqc\!\Big( W^2 \frac{\pa \L }{\pa u}\Big)\right)\right] \\
&+& \frac{\rm i}{2} W^2 {\bar W}^2 (\cD_{\a \ad} {\bar u})
\frac{\pa \L }{\pa {\bar u}}
~-~ \frac{\rm i}{16} W^2 (\cD_{\a \ad} {\bar W}^2)
\Dsqac \Big({\bar W}^2 \, \frac{\pa \L }{\pa {\bar u}} \Big)~. \non
\eea

\subsection{Coupling to new minimal supergravity}

It is known that new minimal supergravity
can be treated
as a super-Weyl invariant dynamical system describing
the coupling of old minimal supergravity  to a {\it real}
covariantly linear scalar superfield ${\Bbb L}$,
\be
\Dsqac\, {\Bbb L} =
(\cD^2 - 4 {\bar R})\, {\Bbb L} = 0~.
\label{linear}
\ee
Any system of matter superfields $\J$ coupled to
new minimal supergravity can be treated as
a super-Weyl invariant coupling
of old minimal supergravity to the matter
superfields $\J$ and $\Bbb L$ (see \cite{BK} for a review).
It is clear that matter couplings in
the new minimal formulation of supergravity
are more restrictive as compared to the old
minimal version. Here we will demonstrate how to couple
the models of self-dual nonlinear electrodynamics to
new minimal supergravity.

Super-Weyl transformations, originally introduced in \cite{HT},
are simply local rescalings of the chiral compensator in old minimal
supergravity (see \cite{GGRS,BK}). In terms of the covariant 
derivatives,
the super-Weyl transformation\footnote{Under
(\ref{superweyl}), the full superspace measure changes as
${\rm d}^8 z\, E^{-1} \to
{\rm d}^8 z\, E^{-1} \,\exp (\s + \bar \s ) $,
while the {\it chiral superspace measure} transforms as
${\rm d}^8 z\, E^{-1} / R \to
{\rm d}^8 z\, (E^{-1} /R ) \, \exp (3\s )$,
see eq. (\ref{chiralrule}).} is
\be
\cD_\a ~\to~ {\rm e}^{ \s/2 - {\bar \s} } \Big(
\cD_\a - (\cD^\b \s) \, M_{\a \b} \Big) ~, \qquad
\cDB_\ad ~\to~ {\rm e}^{ {\bar \s}/2 - \s } \Big(
\cDB_\ad -  (\cDB^\bd {\bar \s}) {\bar M}_{\bd\ad} \Big)~,
\label{superweyl}
\ee
with $\s(z) $ an arbitrary  covariantly chiral scalar parameter,
$\cDB_\ad \s=0$. Since
\be
(\cD^2 - 4 {\bar R}) ~ \to ~{\rm e}^{-2 \bar \s} \,
(\cD^2 - 4 {\bar R})\,{\rm e}^{ \s}
\ee
when acting on a scalar superfield, it is clear
that the super-Weyl transformation law of $\Bbb L$ is
uniquely fixed  to be
\be
{\Bbb L} ~\to ~ {\rm e}^{-\s - \bar \s} \, {\Bbb L}~.
\ee
The new minimal supergravity action\footnote{In the flat
superspace limit, when we set $H^m =0$ and $\vf =1$,
such an action describes the so-called improved tensor multiplet
\cite{dWR}.}
is
\be
S_{\rm SG,new} = 3 \intss\, E^{-1}\,
{\Bbb L}\, {\rm ln} {\Bbb L}~.
\label{nsg}
\ee

Given a massless vector multiplet,
eq. (\ref{eq:w-bar-w}), the gauge field  $V$ is inert
under the super-Weyl transformations, while $W_\a$ changes as
\be
W_\a ~\to ~ {\rm e}^{-3  \s /2} \, W_\a ~.
\ee
It is then clear that only the kinetic term in
(\ref{eq:family action}) is super-Weyl invariant.
However, the nonlinear part of (\ref{eq:family action})
can be promoted to a super-Weyl invariant functional
by coupling the vector multiplet to $\Bbb L$. To achieve this,
it is sufficient to notice that
the combination\footnote{Here we actually generalize
the construction \cite{KT1} of $\cN=1$ superconformal
U(1) duality invariant systems in flat superspace.}
\be
(\cD^2 - 4 {\bar R}) \Big( {W^2 \over {\Bbb L}^2 } \Big)
\ee
is super-Weyl invariant. As a result, we can replace the action
(\ref{eq:family action}) by the following
functional\footnote{Without spoiling the super-Weyl invariance and
self-duality of the action (\ref{SED-NSG}),
the `compensator' $\Bbb L$ can be replaced in (\ref{SED-NSG})
by ${\Bbb L}/ \k$, with $\k$ a coupling constant. We set this
constant to be one since it can be absorbed via
renormalization of the self-interaction,
$\hat{\L}(u, \bar u ) = \k^2 \L (\k^2 u, \k^2 {\bar u})$,
see \cite{KT2}.}
\bea
S[W,{\bar W},{\Bbb L}] =
\frac{1}{4}\intss\, \ERc\, W^2 &+&
\frac{1}{4}\intss\, \ERac\, {\bar  W}^2
\non \\
&+&
\frac14\, \intss\, E^{-1} \,
\frac{W^2\,{\bar W}^2}{{\Bbb L}^2}\,
\L\!\left(\frac{u}{{\Bbb L}^2},
\frac{\bar u}{{\Bbb L}^2}\right)~,
\label{SED-NSG}
\eea
which is (i) super-Weyl invariant and (ii) self-dual, i.e.
it solves the self-duality equation (\ref{eq:duality}).
This action describes self-dual supersymmetric electrodynamics
in new minimal supergravity.

\subsection{Coupling to nonlinear sigma models}\label{sect:sigma}
As is known, supersymmetric nonlinear sigma models are most easily
described in the framework of new minimal supergravity
(see, e.g. \cite{BK} for a review).
Given a K\"ahler manifold parametrized by complex coordinates
$\f$ and their conjugates $\bar \f$, with $K(\f, \bar \f )$
the K\"ahler potential, the corresponding
supergravity-matter action is
\be
S= 3 \intss\, E^{-1}\,
{\Bbb L}\, {\rm ln} {\Bbb L} + \intss\, E^{-1}\,
{\Bbb L}\,K(\f, \bar \f )~.
\label{sigma}
\ee
The dynamical variables $\f$ are covariantly chiral scalar
superfields, $\cDB_\ad \f=0 $, being inert with respect to
the super-Weyl transformations.
The action is obviously super-Weyl invariant.
Moreover, the action is invariant under
the K\"ahler transformations
\be
K(\f, \bar \f) \to K(\f, \bar \f) + \l(\f) + \bar \l(\bar \f)~,
\label{eq:kahler}
\ee
with $\l(\f)$  an arbitrary holomorphic function.

It is easy to couple the above model to the self-dual
supersymmetric electrodynamics (\ref{SED-NSG}).
The supergravity-matter system is described by
the action
\be
S[W,{\bar W}, \f, {\bar \f},{\Bbb L}]
= 3 \intss\, E^{-1}\,
{\Bbb L}\, {\rm ln} {\Bbb L} + \intss\, E^{-1}\,
{\Bbb L}\,K(\f, \bar \f )
+ S[W,{\bar W},{\Bbb L}] ~,
\label{NSG-ED-sigma}
\ee
and this theory possesses several important symmetries:
(i) super-Weyl invariance; (ii) K\"ahler invariance;
(iii) U(1) duality invariance.
We should now uncover the description of this theory
in the framework of old minimal supergravity.

Let us replace the action (\ref{NSG-ED-sigma})
by the following auxiliary action
\be
S[W, {\bar W}, \f, {\bar \f}, {\Bbb L}, U] =
3 \intss\, E^{-1} \left(U\, {\Bbb L} - \U\right)
+ S[W,{\bar W}, \U] ~,
\label{eq:auxiliary}
\ee
where
\be
\U  = {\rm exp}\!\left(U - \frac13
K(\f, {\bar \f})\right)~.
\label{eq:upsilon}
\ee
Here the additional dynamical variable $U$
is an unconstrained real scalar
superfield, and the action $S[W,{\bar W}, \U] $ is obtained from
(\ref{SED-NSG}) by replacing ${\Bbb L} \to \U$.
In order for the action to be super-Weyl invariant, the superfield
$U$ must possess the following super-Weyl transformation law:
\be
U ~\to ~ U - \s - {\bar \s} ~.
\label{U-super-Weyl}
\ee

The theories (\ref{NSG-ED-sigma}) and (\ref{eq:auxiliary})
are equivalent to each other. Indeed, the $U$-equation of motion
derived from (\ref{eq:auxiliary}) is algebraic and it can be uniquely
solved by expressing $U$ in terms of the other superfields.
Upon elimination of $U$ in this way we regain the action
(\ref{NSG-ED-sigma}). On the other hand, let us consider the
$\Bbb L$-equation of motion derived from (\ref{eq:auxiliary}):
$\Dsqac \cD_\a\,U=0$. The general solution to this
equation is just the requirement that $U$
be the sum of a covariantly chiral scalar
superfield and its conjugate; as a result, the linear superfield
$\Bbb L$ completely decouples.
Now, the super-Weyl gauge freedom (\ref{U-super-Weyl})
allows us to impose the gauge condition $U=0$, and the action
(\ref{eq:auxiliary}) then becomes
\be
S[W, {\bar W}, \f, {\bar \f}] =
-3 \intss\, E^{-1} \,
{\rm e}^{-{1 \over 3} K(\f, \bar \f )}
+ S[W,{\bar W}, {\rm e}^{ -{1 \over 3} K(\f, \bar \f ) }] ~.
\label{OSG-ED-sigma}
\ee
This is the old minimal supergravity counterpart
of the model (\ref{NSG-ED-sigma}).
To preserve the super-Weyl gauge condition
$U=0$, any K\"ahler transformation (\ref{eq:kahler})
must now be accompanied by the induced super-Weyl
transformation with $\s = {1 \over 3}\l(\f)$.
As a result,  one ends up with the so-called
super-Weyl--K\"ahler transformations
(see \cite{GGRS,WB,BK} for more details).

\sect{Coupling to the dilaton-axion multiplet}
\label{sect:matter}

In the above analysis of the coupling of self-dual supersymmetric
electrodynamics to K\"ahler sigma models in curved superspace,
it was assumed that the matter superfields, $\f$ and $\bar \f$,
are inert under the electromagnetic duality rotations.
Of some interest is a more general situation
when, say,  a chiral matter superfield $\F$ and its conjugate $\bar \F$
{\it do} transform under duality rotations, which can now span
a larger group than the one corresponding to the pure gauge
field case. Coupling to the so-called dilaton-axion supermultiplet
is an important example.

We start by formulating the conditions of duality invariance
for the Abelian vector multiplet
$(W_\a, \,{\bar W}_\ad )$ interacting with chiral matter
$(\F, \,{\bar \F})$ in curved superspace.
Let $S[v] = S[W, {\bar W}, \F,{\bar \F}]$
be the action functional of the supergravity-matter systems,
with the dependence of $S[v]$ on the supergravity prepotentials
being implicit.
We again introduce covariantly  (anti) chiral spinor superfields
${\bar M}^{\ad }$ and $M_\a$ defined
by the rule (\ref{eq:Mdefinition}).
Since the Bianchi
identity $\cD^\a W_\a = \cDB_\ad {\bar W}^{\ad }$ and the
gauge field equation of motion $\cD^\a M_\a = \cDB_\ad {\bar M}^{\ad }$
are of the same functional form, we may consider
infinitesimal duality transformations
\bea
\d \left( \begin{array}{c} M_\a  \\  W_\a  \end{array} \right)
&=&  \left( \begin{array}{cc} ~a~& ~b~ \\
~c~ & ~d~ \end{array} \right)
\left( \begin{array}{c} M_\a \\ W_\a  \end{array} \right) ~, \qquad 
\quad
\d \F = \x (\F)~,
\label{eq:matter duality transformation}
\eea
with $\x (\F )$ a {\it holomorphic} function
and $a, b, c$ and $d$ real numbers.

${}$Following \cite{KT2},
the conditions of duality invariance in the presence of matter
can be shown to be
\bea
\d \F \cdot \frac{\d S}{\d \F } +
\d {\bar \F} \cdot \frac{\d S}
{\d {\bar\F} }
&=& \frac{\rm i}{4} \, b \,\Big( W \cdot W -
{\bar W} \cdot {\bar W} \Big) - \frac{\rm i}{4}\, c \,\Big(
M \cdot M -
{\bar M} \cdot {\bar M} \Big) \non \\
&+& \frac{\rm i}{2} \,a  \,\Big(
W \cdot M -
{\bar W} \cdot {\bar M} \Big)~,
\label{eq:sd matter equations}
\eea
with $d=-a$.
We see that the  maximal group of duality transformations is
${\rm Sp}(2, \dsR) \cong {\rm SL}(2, \dsR)$.
The complex variable $\F$ should then parametrize the
homogeneous space SL$(2,{\Bbb R}) /$U(1), with the vector
field $\x (\F)$ in (\ref{eq:sd matter equations})
generating the action of SL$(2,{\Bbb R})$ on the coset space.
The matter-free case, which was considered before,
corresponds to freezing the superfield $\F(z)$ to a given
point of the space SL$(2,{\Bbb R}) /$U(1).  In such a case,
the duality group, SL$(2,{\Bbb R})$,  reduces to U(1) --
the stabilizer of the point chosen.

To describe the dilaton-axion multiplet,
we make use of the lower half-plane
realization of the coset space SL$(2,{\Bbb R}) /$U(1).
Then, the variation $\d \F = \x (\F)$ in
(\ref{eq:matter duality transformation}) is
\be
\d \F = b +2a \,\F -c \, \F^2~.
\ee
Our solution to the equations (\ref{eq:sd matter equations}) reads
\bea
S &=& 3 \intss\, E^{-1}\,
{\Bbb L}\, {\rm ln} {\Bbb L} + \intss\, E^{-1}\,
{\Bbb L}\, \Big( \cK(\F, \bar \F ) + K(\f, \bar \f) \Big) \non \\
&&
+ \frac{\rm i}{4}\intss\, \ERc\, \F\,W^2 -
\frac{\rm i}{4}\intss\, \ERac\,
{\bar \F}\,{\bar  W}^2
\label{eq:dil-ax1}
\\
&&-  \frac{1}{16}\, \intss\, E^{-1} \,
(\F-{\bar \F})^2\, \frac{W^2\,{\bar W}^2}{{\Bbb L}^2}  \,
\L \Big( \frac{\rm i}{2} (\F - {\bar \F}) \,
{u \over {\Bbb L}^2 } \; , \;
\frac{\rm i}{2} (\F-{\bar \F}) \,
{ {\bar u} \over {\Bbb L}^2 } \Big)~,
\non
\eea
and $u$ is defined in (\ref{u}). Here $\cK(\F, \bar \F ) $
is the K\"ahler potential of the K\"ahler manifold
SL$(2,{\Bbb R}) /$U(1).  The term
$\intss\, E^{-1}\,{\Bbb L}\,K(\f, \bar \f) $ in (\ref{eq:dil-ax1})
   corresponds
to the chiral matter which is inert under the duality rotations.
For $\F= - {\rm i}$, the action (\ref{eq:dil-ax1})
reduces to (\ref{NSG-ED-sigma}).

The supergravity-matter system (\ref{eq:dil-ax1}) enjoys
the following important properties:
(i) super-Weyl invariance; (ii) K\"ahler invariance;
(iii)  SL$(2,{\Bbb R})$ duality invariance.
To re-formulate this theory in the framework of the old minimal
version of $\cN=1$ supergravity, one should eliminate
the real linear compensator $\Bbb L$
following the  procedure described in subsection \ref{sect:sigma}.
This will lead to
\bea
S &=& -3 \intss\, E^{-1}\,
\exp  \!\Big( -{1 \over 3}\cK(\F, \bar \F )
-{1 \over 3} K(\f, \bar \f) \Big)
\non \\
&&+ \frac{\rm i}{4}\intss\, \ERc\, \F\,W^2 -
\frac{\rm i}{4}\intss\, \ERac\,
{\bar \F}\,{\bar  W}^2
\\
&&-  \frac{1}{16}\, \intss\, E^{-1} \,
(\F-{\bar \F})^2\, \frac{W^2\,{\bar W}^2}{ {\bf \U}^2 }  \,
\L \Big( \frac{\rm i}{2} (\F-{\bar \F}) \,{ u \over {\bf \U}^2} \;, \;
\frac{\rm i}{2} (\F-{\bar \F}) \,{{\bar u} \over {\bf \U}^2 } \Big)~,
\non
\eea
where
\be
{\bf \U} = \exp   \! \Big(-{1 \over 3}\cK(\F, \bar \F )
-{1 \over 3} K(\f, \bar \f) \Big) ~.
\ee
Unlike (\ref{eq:dil-ax1}), this action enjoys
the super-Weyl--K\"ahler invariance.

To describe the dilaton-axion complex, we have used
the $\cN=1$ chiral multiplet. In the context of heterotic string
theory, the dilaton-axion complex is realized in terms of the
$\cN=1$ tensor multiplet. Transition from the chiral
to the tensor realization can be implemented as follows.
The dilaton-axion K\"ahler potential can be chosen to be
$\cK(\F, \bar \F ) = - \k^2 \ln \,
{\rm i} (\F -    \bar \F )$, with $\k$ a constant.
As a result, the action  (\ref{eq:dil-ax1}) can be brought
(at the cost of sacrificing the manifest gauge invariance in the
second line of the action)
to such a form that  $\F$ and $\bar \F$ appear only in the real
combination ${\rm i} (\F -    \bar \F )$. We can then apply
a superfield Legendre transformation which turns
the description in terms of $\F$ and $\bar \F$
into the one in terms of a {\it real} superfield $\Bbb G$
under the modified linearity condition
\be
(\cDB^2 - 4R) {\Bbb G} = W^\a W_\a~, \qquad
(\cD^2 - 4{\bar R}) {\Bbb G} = {\bar W}_\ad {\bar W}^\ad~.
\ee
This constraint is known to describe the Chern-Simons
coupling of the tensor multiplet to the vector multiplet.
An interesting open question is: What is the fate of the
SL$(2,{\Bbb R})$ duality symmetry in this dual version of the theory?

\vskip.5cm

\noindent
{\bf Acknowledgements.}
We are grateful to Ian McArthur for discussions and for reading
the manuscript. The work of SK  is supported by the Australian Research
Council, the Australian Academy of Science as well as by UWA research 
grants.
The work of SMc is supported by the Hackett Postgraduate Scholarship.


\begin{thebibliography}{99}

\bi{Sch} E. Schr\"odinger, Proc.\ Roy.\ Soc.\ {\bf A150}
(1935) 465.

\bi{BI}
M.~Born and L.~Infeld, Proc. Roy. Soc.
{\bf A144} (1934) 425.

\bi{FT} E.S.~Fradkin and A.A.~Tseytlin,
Phys. Lett. {\bf B163} (1985) 123.

\bi{L} R.G.~Leigh, Mod. Phys. Lett. {\bf A4}
(1989) 2767.

\bibitem{FSZ}
S.~Ferrara, J.~Scherk and B.~Zumino,
Nucl.\ Phys.\  {\bf B121} (1977) 393.

\bibitem{CJ}
E.~Cremmer and B.~Julia,
Nucl.\ Phys.\  {\bf B159} (1979) 141.

\bi{GZ1}
M.K.~Gaillard and B.~Zumino,
Nucl.\ Phys.\  {\bf B193} (1981) 221

\bi{Z}
B.~Zumino,
in {\it Quantum
Structure of Space and Time}, M.~J. Duff and
C.~J. Isham (Eds.), Cambridge University Press, 1982 p. 363.

\bibitem{GR1}
G.W.~Gibbons and D.A.~Rasheed,
Nucl.\ Phys.\  {\bf B454} (1995) 185
[hep-th/9506035].

\bibitem{GR2}
G.W.~Gibbons and D.A.~Rasheed,
Phys.\ Lett.\  {\bf B365} (1996) 46
[hep-th/9509141].

\bi{GZ2}
M.K.~Gaillard and B.~Zumino,
in {\it Supersymmetry and Quantum Field Theory},
J.~Wess and V.P.~Akulov (Eds.), Springer Verlag, 1998,
p. 121
[hep-th/9705226].

\bi{GZ3}
M.K.~Gaillard and B.~Zumino,
in {\it Duality and
Supersymmetric Theories}, D.I.~Olive and
P.C.~West (Eds.), Cambridge University Press,
1999, p. 33 [hep-th/9712103].

\bibitem{list}
Y.~Tanii,
{\it Introduction to supergravities in diverse dimensions},
hep-th/9802138;
M.~Araki and Y.~Tanii,
Int.\ J.\ Mod.\ Phys.\  {\bf A14} (1999) 1139
[hep-th/9808029];
T.~Kimura and I.~Oda,
Int.\ J.\ Mod.\ Phys.\  {\bf A16} (2001) 503
[hep-th/9904019];
D.~Brace, B.~Morariu and B.~Zumino,
in {\it The Many Faces of the Superworld: Yury Golfand
Memorial Volume}, M. Shifman (Ed.),
World Scientific, 2000, p. 103 [hep-th/9905218];
P.~Aschieri, D.~Brace, B.~Morariu and B.~Zumino,
Nucl.\ Phys.\  {\bf B574} (2000) 551
[hep-th/9909021];
M.~Hatsuda, K.~Kamimura and S.~Sekiya,
Nucl.\ Phys.\  {\bf B561} (1999) 341
[hep-th/9906103];
P.~Aschieri, D.~Brace, B.~Morariu and B.~Zumino,
Nucl.\ Phys.\ {\bf B588} (2000) 521
[hep-th/0003228].

\bibitem{GH}
G.W.~Gibbons and K.~Hashimoto,
JHEP {\bf 0009} (2000) 013
[arXiv:hep-th/0007019].

\bibitem{KT1}
S.M.~Kuzenko and S.~Theisen,
JHEP {\bf 0003} (2000) 034 [hep-th/0001068].

\bibitem{KT2}
S.M.~Kuzenko and S.~Theisen,
Fortsch.\ Phys.\  {\bf 49} (2001) 273 [hep-th/0007231].

\bibitem{IZ}
E.A.~Ivanov and B.M.~Zupnik,
Nucl.\ Phys.\ {\bf B618} (2001) 3 [hep-th/0110074];
{\it New representation for Lagrangians of self-dual nonlinear
electrodynamics}, 
hep-th/0202203.

\bibitem{BG1}
J.~Bagger and A.~Galperin,
Phys.\ Rev.\ {\bf D55} (1997) 1091 [hep-th/9608177].

\bibitem{BG2}
J.~Bagger and A.~Galperin,
Phys.\ Lett.\  {\bf B412} (1997) 296
[hep-th/9707061].

\bi{RT}
M.~Ro\v{c}ek and A.A.~Tseytlin,
Phys.\ Rev.\ {\bf D59} (1999) 106001
[hep-th/9811232].

\bibitem{CF} S.~Cecotti and S.~Ferrara, Phys. Lett.
{\bf B187} (1987) 335.

\bibitem{Ket}
S.V.~Ketov,
Nucl.\ Phys.\ {\bf B553} (1999) 250 [hep-th/9812051];
Class.\ Quant.\ Grav.\  {\bf 17} (2000) L91 [hep-th/0005126].

\bibitem{BIK}
S.~Bellucci, E.~Ivanov and S.~Krivonos,
Phys.\ Lett.\ {\bf B502} (2001) 279 [hep-th/0012236];
Phys.\ Rev.\ {\bf D64} (2001) 025014 [hep-th/0101195];
Nucl.\ Phys.\ Proc.\ Suppl.\  {\bf 102} (2001) 26 [hep-th/0103136].

\bibitem{GKPR}
F.~Gonzalez-Rey, B.~Kulik, I.~Y.~Park and M.~Ro\v{c}ek,
Nucl.\ Phys.\ {\bf B544} (1999) 218 [hep-th/9810152].

\bi{WZ-old}
J.~Wess and B.~Zumino,
Phys.\ Lett.\ {\bf B66} (1977) 361;
R.~Grimm, J.~Wess and B.~Zumino,
Phys.\ Lett.\ {\bf B73} (1978) 415;
J.~Wess and B.~Zumino,
Phys.\ Lett.\ {\bf B74} (1978) 51.

\bibitem{old}
K.S.~Stelle and P.C.~West,
Phys.\ Lett.\ {\bf B74} (1978) 330;
S.~Ferrara and P.~van Nieuwenhuizen,
Phys.\ Lett.\ {\bf B74} (1978) 333.

\bibitem{new}
V.P.~Akulov, D.V.~Volkov and V.A.~Soroka,
Teor.\ Mat.\ Fiz.\  {\bf 31} (1977) 12;
M.F.~Sohnius and P.C.~West,
Phys.\ Lett.\ {\bf B105} (1981) 353.

\bi{non-min}
P.~Breitenlohner,
Nucl.\ Phys.\ {\bf B124} (1977) 500;
W.~Siegel and S.J.~Gates,
Nucl.\ Phys.\ {\bf B147} (1979) 77.

\bibitem{GGRS}
S.J.~Gates, M.T.~Grisaru, M.~Ro\v{c}ek and W.~Siegel,
{\it Superspace, or One Thousand and One Lessons in Supersymmetry},
Front.\ Phys.\  {\bf 58} (1983) 1 [hep-th/0108200].

\bibitem{WB} J.~Wess and J.~Bagger,
{\it Supersymmetry and Supergravity},
Princeton Univ. Press, 1992.

\bibitem{BK} I.L.~Buchbinder and S.M.~Kuzenko,
{\it Ideas and Methods of Supersymmetry and
Supergravity or a Walk Through Superspace},
IOP, Bristol, 1998.

\bibitem{GK}
S.J.~Gates and S.V.~Ketov,
Class.\ Quant.\ Grav.\ {\bf 18} (2001) 3561 [hep-th/0104223].

\bibitem{HT}
P.S.~Howe and R.W.~Tucker,
Phys.\ Lett.\ {\bf B80} (1978) 138.

\bibitem{dWR}
B.~de Wit and M.~Ro\v{c}ek,
Phys.\ Lett.\ {\bf B109} (1982) 439.

\end{thebibliography}
\end{document}